\documentclass[preprint,showpacs,preprintnumbers,amsmath,amssymb]{revtex4}

\usepackage{graphicx}
\usepackage{dcolumn}
\usepackage{bm}

\begin{document}

\preprint{}

\title{Local stability conditions for asymmetric Fermi superfluid}%
\author{S.-K. Yip}
\affiliation{ Institute of Physics, Academia
Sinica, Nankang, Taipei 115, Taiwan}%

\date{\today}

\begin{abstract}
Two different
local stability conditions for an asymmetric
superfluid has been discussed in the literature.
We here consider the relations between them.

\end{abstract}

\maketitle

In our recent paper \cite{UNpd}, we considered a two
component Fermi gas with unequal populations
under a short-ranged attractive interaction.
We constructed the phase diagram by first finding
uniform state solutions to the number and gap equations.
We then impose the local \cite{local} stability condition
that the susceptibility matrix has to have only
positive eigenvalues.   In particular then, in terms
of the differences in populations and chemical
potentials $n_d$ and $h$, 
local stability/instability is decided via
\begin{equation}
\left( \frac{ \partial n_d} {\partial h} \right)_{\mu} 
{> \atop <} 0
\label{sus}
\end{equation}
Here $\mu$ is the average chemical potential. \cite{cut}
  There we stated, without providing details,
that a negative susceptibility implies instability
towards phase separation.  We thus then exclude all
states with negative susceptibilities and identified
the corresponding point in the phase diagram as not
having a stable uniform state.

The phase diagram and local stability condition have
been discussed in a different manner 
\cite{Sarma,Bedaque,SRn,SRw,SRcom,Parish,Gu}.  
These authors considered
the free energy of the system 
$\tilde \Omega (\Delta; \mu, h)$ as a function of the order parameter
$\Delta$ for fixed $\mu$ and $h$ \cite{tilde}.  The locally stable
states are selected according to the condition that
$\tilde \Omega$ must be a relative minimum of $\tilde \Omega$
as a function of $\Delta$.  Thus, at the extremum points
$\left(
\frac{\partial \tilde \Omega}{\partial \Delta} \right)_{\mu,h} = 0$,
they apply the condition
\begin{equation}
\left(
\frac{\partial^2 \tilde \Omega}{\partial \Delta^2} 
  \right)_{\mu,h}  
{> \atop <} 0
\label{dOdh}
\end{equation}
to distinguish locally stable and unstable states.

Since our condition eq (\ref{sus}) was stated without
providing detailed justifications,  there
are some confusions as to how the conditions
(\ref{sus}) and (\ref{dOdh}) are related.
Sheehy and Radzihovsky \cite{SRw,SRcom}
suggest recently that states with 
positive susceptibilities can still correspond
to a relative maximum of the free energy 
$\tilde \Omega (\Delta; \mu, h)$.
Other groups \cite{He06,Chien,Chen2} have set out
to show the relation between the conditions
(\ref{sus}) and (\ref{dOdh}).  
These proofs, while correct, involve algebra 
which motivation 
may be {\it a priori} somewhat mysterious.
The purpose of this note is to provide perhaps
a physically more transparent picture on the relations between
the conditions (\ref{sus}) and (\ref{dOdh}).

\begin{figure}
\includegraphics{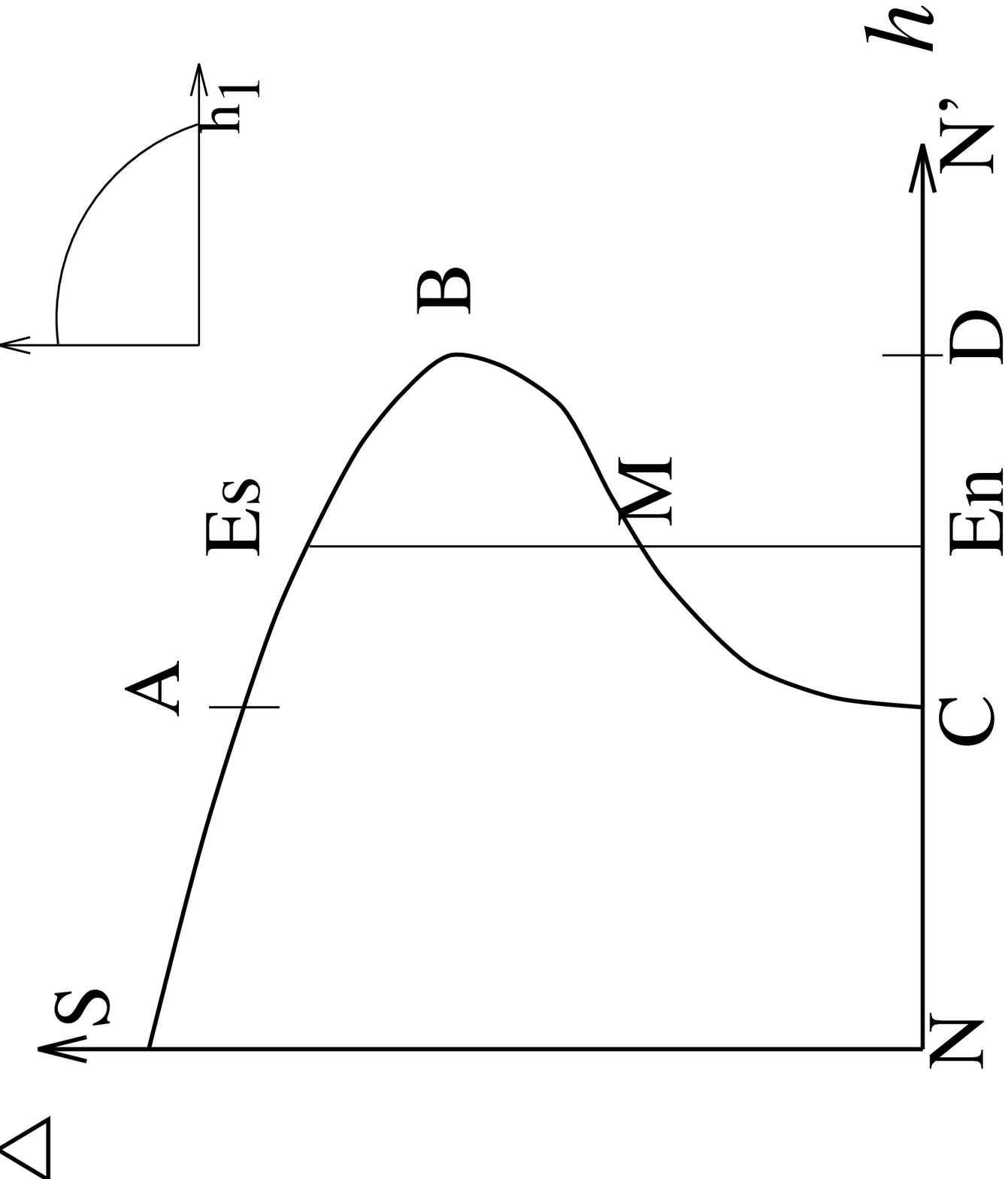} \vspace*{3 in}
 \caption{ $\Delta$ versus $h$ at a given chemical potential.
\label{fig:Dh} }
\end{figure}

We start by considering the locus of the relative extremum
of $\tilde \Omega (\Delta, \mu, h)$, that is,
the solution to the gap equation
$ \left( \frac{\partial \tilde \Omega}{ \partial \Delta} \right)_{\mu,h}
  = 0$.  This gives $\Delta$ as a function of $\mu$ and $h$.
For definiteness \cite{cut}, we consider a "cut" of this function at some
given $\mu$.  First, let us note that the normal state $\Delta = 0$ is
always a solution to this gap equation, and hence
always an extremum of $\tilde \Omega$ (when considered a 
function of $\Delta$ at fixed $\mu$ and $h$).
For some region of parameters, there may be in addition
non-trivial ($\Delta \ne 0$) solution to the gap equation.
An example deep in the BEC (strong attractive interaction)
limit is as sketched in
the inset of Fig \ref{fig:Dh}.
The non-trivial solution to the gap equation gives
a $\Delta$ that is monotonically decreasing with $h$. 
(This non-trivial solution for $h < h_1$ 
 represents a gapless superfluid \cite{UNpd}).
This function however is not necessarily single valued.
An example is as shown in the main part of Fig \ref{fig:Dh}.
This happens in particular in the weak interaction limit,
as was first shown by Sarma \cite{Sarma} (\cite{zeroT})
(and continues to hold until the attractive
interaction is sufficiently strong).  
There, the non-trivial solution for $\Delta$,
instead of decreasing monotonically with $h$,
 "bends around" at the point B.
For $h< h_A$, there is one non-trivial solution 
(labelled by S-A), whereas there are two non-trivial solutions
for $h_A < h < h_B$ (labelled by A-B and B-C).

First we note a simple but useful fact.  The free energy
$\tilde \Omega$ must continue to increase as $\Delta \to \infty$,
as $\Delta = \infty$ is not a solution to the gap equation,
and $\tilde \Omega$ cannot decrease indefinitely.
Since the solutions to the gap equation already give
all the extrema, it follows that the 
solution to the gap equation with the largest $\Delta$ always
corresponds to a relative minimum of $\tilde \Omega$.
In fact, we have the following statements:

\noindent (1) If $\Delta=0$ is the only solution to the gap equation,
then it is the free energy minimum.

\noindent (2) If there is only one non-trivial solution $\Delta = \Delta_1$,
then it is the free energy minimum, and the normal state
is a relative maximum. (In this case, starting from $\Delta=0$,
$\tilde \Omega$ first decreases, then reaches its minimum at 
$\Delta_1$, then increases for all $\Delta > \Delta_1$.)
This applies to the inset for $h < h_1$, and $h< h_A$
for the main Fig \ref{fig:Dh}.

\noindent  (3) If there are two non-trivial solutions $\Delta_1 < \Delta_2$,
then $\Delta_1$ must be a relative maximum, 
whereas both $\Delta_2$ and the normal state
are relative minima. 
(In this case, starting from $\Delta=0$,
$\tilde \Omega$ first increases, then reaches a relative maximum at 
$\Delta_1$, decreases till $\Delta_2$,
then increases again for all $\Delta > \Delta_2$.

Therefore, for the non-trivial solutions in Fig \ref{fig:Dh},
the branch S-A is the absolute minimum for $\tilde \Omega$,
A-B is a relative minimum, B-C must be a
relative maximum.  For the trivial solution $\Delta=0$,
it corresponds to a relative maximum for $h < h_C = h_A$
and a relative minimum for $h > h_C$.
(see also the last paragraph of this note).

\begin{figure}
\includegraphics{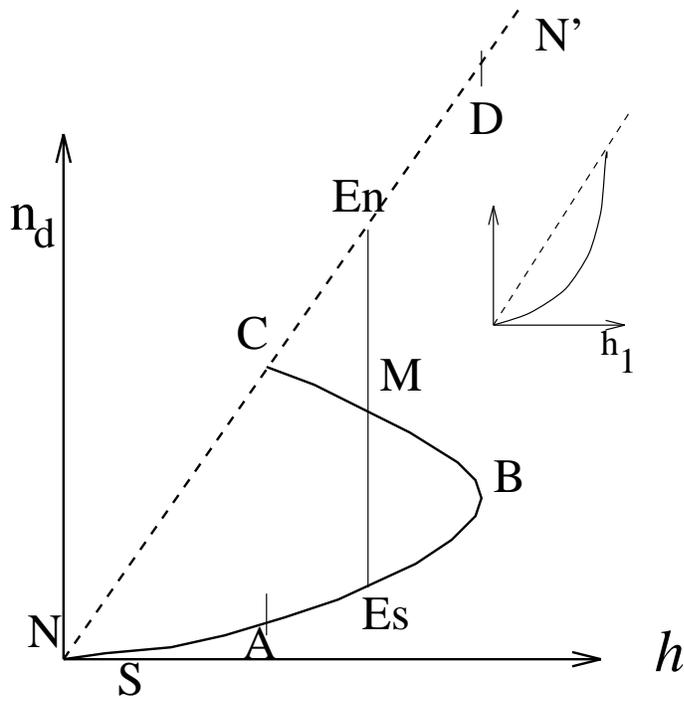} \vspace*{4 in}
 \caption{ $n_d$ versus $h$ at a given chemical potential.
\label{fig:nh} }
\end{figure}

Now we consider $n_d$ as a function of $h$.
For the situation in the inset of Fig \ref{fig:Dh},
there is one branch of $n_d$ for the normal state,
and one for the superfluid state.
A sketch is as shown in the inset of Fig \ref{fig:nh},
where these two branches are denoted by
the dashed and full lines respectively.
Note that the full line must have positive slope as it
is stable. 
Therefore there is one possible $n_d$ for each h for $h> h_1$,
but two for each $h < h_1$.
Now consider decreasing the attractive interaction
(or increasing $\mu$)
so that the $\Delta$-$h$ plot evolves from the form in the inset
of Fig \ref{fig:Dh} to that in the main part of that figure.
Consider now what must happen to the $n_d$-$h$ plot.
There continues to be one branch (dashed line in main part
of Fig \ref{fig:nh}) which corresponds to the normal state.
For the non-trivial solution to the gap equation (S-A-B-C in 
Fig \ref{fig:Dh}), there is an S-A-B part in
Fig \ref{fig:nh} with positive slope corresponding
to the locally stable S-A-B part in Fig \ref{fig:Dh}.
However, since
there is one $n_d$ for each solution to the gap equation, 
there must be two possible $n_d$'s corresponding to
the non-trivial solutions in the range
 $h_C= h_A < h < h_B = h_D$.  
Thus if we follow the path S-A-B-C in Fig \ref{fig:Dh},
the corresponding path in the $n_d$ versus $h$ plot 
must also "turn around"" at point B,
where the $\Delta$ versus $h$ curve turns around.  At this point,
$\left( \frac{\partial n_d}{\partial h} \right)_{\mu}$
necessarily diverges and changes sign at B,
As a result,
the branch BC in Fig \ref{fig:nh} has
$\left( \frac{\partial n_d}{\partial h} \right)_{\mu} < 0$,
corresponding to 
$\left(
\frac{\partial^2 \tilde \Omega}{\partial \Delta^2} 
  \right)_{\mu,h}  < 0$ in Fig \ref{fig:Dh}.

At this point let us make connection with
the arguments given in \cite{He06,Chien,Chen2,Reply}.
In essence, these arguments are based on
the following equation or its variation:

\begin{equation}
\left( \frac{ \partial n_{d}}{\partial h} \right)_{\mu}
=
\left( \frac{ \partial \tilde n_{d}}{\partial h} \right)_{\Delta,\mu}
+
\frac
{ \left( \frac{\partial \tilde n_d}{\partial \Delta} 
   \right)_{\mu,h}^2 }
{ \left(
\frac{\partial^2 \tilde \Omega}{\partial \Delta^2}
  \right)_ {\mu,h}}
\label{matrix}
\end{equation}
Here
$\tilde n_d$ is a function of $\Delta$, $\mu$, $h$
defined by the relation 
$\tilde n_d \equiv - \left( \frac{ \partial \tilde \Omega}
{\partial h} \right)_{\Delta,\mu}$.
Note that, at the solution to the gap equation,
$n_d (\mu,h) = \tilde n_d (\Delta, \mu, h)$
since  
$n_d =  - \left( \frac{ \partial \Omega}
{\partial h} \right)_{\mu} 
 = - \left( \frac{ \partial \tilde \Omega}
{\partial h} \right)_{\Delta,\mu}$
since $ \left( \frac{ \partial \tilde \Omega} 
        {\partial \Delta} \right)_{\mu,h} = 0$.
Suppose we follow the line S-A-B-C in Fig \ref{fig:Dh}.
At B, one changes from a branch
where
$\left(
\frac{\partial^2 \tilde \Omega}{\partial \Delta^2}
  \right)_ {\mu,h} > 0$ to $< 0$.
Since
$\left(
\frac{\partial^2 \tilde \Omega}{\partial \Delta^2}
  \right)_ {\mu,h}$ appears in the denominator
of eq (\ref{matrix}), 
$\left( \frac{ \partial n_{d}}{\partial h} \right)_{\mu}$
must changes from positive to negative through
$\pm \infty$ near B \cite{nonzero}.  
As we have already seen above, this behavior is
a natural consequence of the fact that the curve $n_d$ versus $h$
must turn around at point B in Fig \ref{fig:nh}.

As seen from eq (\ref{matrix}) however,
since 
$\left( \frac{ \partial \tilde n_{d}}{\partial h} \right)_{\Delta,\mu}
> 0$ always (c.f. \cite{SRcom}), there seems no
general principle requiring that
$\left( \frac{ \partial n_{d}}{\partial h} \right)_{\mu}
< 0$ whenever 
$\left( \frac{\partial^2 \tilde \Omega}{\partial \Delta^2}
  \right)_ {\mu,h} < 0$.
Similarly, 
in principle with further change of interaction
parameters, some part of the BC curve in Fig \ref{fig:nh}
may evolve further so that it can acquire a positive slope.
However, for this to happen,
it can easily be seen that one must have some region
of $n_d$ which corresponds to at least three possible $h$'s,
and with at least one point where 
$\left( \frac{\partial n_d}{\partial h}\right)_{\mu}$ vanishes.
This does not seem to happen for the 
current system of asymmetric Fermi gas in mean-field theory.
Moreover,  there is no reason why the change to positive
slope would occur at the same point in parameter space
as where $\Delta$ versus $h$ turns around in Fig \ref{fig:Dh}.
Thus evaluating the susceptibility
should still indicate when one goes from a local
minimum to maximum of $\tilde \Omega$ as a function of $\Delta$.

Finally, we notice here also that Fig \ref{fig:nh} is similiar
to the one in theory of liquid gas transition 
such as the Van der Waals equation of state for a
non-ideal gas.  $h$ and $-n_d$ here plays the role of the
pressure $P$ and volume $V$ respectively.
(A sign change in $n_d$ is necessary since here
$d \Omega = - n_d dh$, whereas for the gas,
the differential of Gibbs free energy is 
$d G = V dP$).  Indeed, Fig \ref{fig:nh}
has precisely that form if we make the above identification
(and ignore the branch NC, which is never stable).
The branch BC there is locally unstable since
the compressibility is negative.
By again the analogy to the Van der Waals gas,
the {\em absolutely} stable branch can be
determined by Maxwell construction.
These branches are S-A-E$_s$ 
for $h< h_E \equiv h_{Es} = h_{En}$ and
 E$_n$-D-N' if $h > h_E$, with
the condition that the areas
E$_s$-B-M and M-C-E$_n$ are equal.
At $h= h_E$, the system phase separates into
a mixture with part of the volume in state E$_s$ and
part in E$_n$.

This research was supported by the NSC of
Taiwan under grant number NSC94-2112-M-001-002.

\newpage


\vskip 1 in


\begin{thebibliography}{99}
\bibitem{local} We shall first discuss
the local stability condition. After the local minima
are found, the state that is globally stable can
be found by comparing the free energies of
these local minima (see below)
\bibitem{tilde} We use a tilde to denote the free
energy as a function of the variable $\Delta$
before it has been eliminated.
The true free energy $\Omega (\mu, h)$
is obtained by evaluating this function
$\tilde \Omega$ at its minimum.
\bibitem{zeroT} 
At zero temperature, part or all of
the branch S-A-B corresponding to the superfluid in
in Figs \ref{fig:Dh} and \ref{fig:nh} have zero slope.
Moreover, strictly speaking the dashed line in Fig \ref{fig:nh} is not
a simple straight line.  
We ignore these complications since they
have no effect on our arguments below.
\bibitem{nonzero}  For this argument to hold,
the quantity $ \left( \frac{\partial \tilde n_d}{\partial \Delta} 
   \right)_{\mu,h}^2 $ must not vanish.
In particular it does not apply to the branch NC.
On approaching C,  one can show
by a Ginzburg-Landau type of argument that
the second term in eq (\ref{matrix}) remains finite.
\end{thebibliography}
\end{document}